\documentclass[aps,prl,floatfix,
twocolumn,
superscriptaddress]{revtex4-2}

\usepackage{amsmath,amssymb}
\usepackage{graphicx}
\usepackage{hyperref}
\usepackage{epsfig}
\usepackage{bm}
\usepackage[usenames]{color}

\usepackage{booktabs}

\begin{document}

\title{Quantum tribology: acceleration-induced Stokes friction and Magnus force\\ in correlated Bose fluids}






\author{V.~M.~Kovalev}
\affiliation{Guangdong Technion -- Israel Institute of Technology, 241 Daxue Road, Shantou, Guangdong, China, 515063}
\affiliation{Rzhanov Institute of Semiconductor Physics, Siberian Branch\\ of Russian Academy of Science, Novosibirsk 630090, Russia}
\affiliation{Novosibirsk State Technical University, Novosibirsk 630073, Russia}

\author{A.~N.~Osipov}
\affiliation{Technion -- Israel Institute of Technology, Haifa, 3200003, Israel}
\affiliation{Guangdong Technion -- Israel Institute of Technology, 241 Daxue Road, Shantou, Guangdong, China, 515063}

\author{O.~V.~Kibis}
\affiliation{Novosibirsk State Technical University, Novosibirsk 630073, Russia}

\author{I.~G.~Savenko}
\email[ivan.g.savenko@gmail.com]{}
\affiliation{Guangdong Technion -- Israel Institute of Technology, 241 Daxue Road, Shantou, Guangdong, China, 515063}
\affiliation{Technion -- Israel Institute of Technology, Haifa, 3200003, Israel}

\date{\today}

\begin{abstract}
The Landau criterion, a cornerstone of quantum fluid dynamics, dictates that dissipation is forbidden for uniform motion below a critical velocity. 
Yet, the fundamental question of how acceleration reshapes the principles of quantum friction has remained open since Landau and Pitaevskii's seminal works. 
Here, we establish a theoretical framework for the quantum tribology of non-inertial motion, describing a probe particle undergoing composite translation and rotation within a weakly interacting Bose condensate. 
Using the nonlinear Gross-Pitaevskii equation, we show that centripetal acceleration fundamentally modifies the energy-momentum constraints on elementary excitations. 
This leads to a finite drag force in the subsonic regime of the probe particle motion, and a characteristic quantum stick-slip behaviour in the deeply supersonic regime -- a direct generalization of the classical Landau-Pitaevskii picture. 
Beyond this dissipative response, we uncover a fundamentally distinct mechanism: the nonlinearity of the quantum fluid, combined with the broken symmetry of the trajectory, gives rise to a non-dissipative anomalous transverse force. 
This quantum Magnus-like response, emerging from the second-order density perturbation, performs no work and is rooted in the geometric asymmetry of the dynamically induced flow. 
Our findings lay the foundation for a universal program in quantum tribology of accelerated motion, establishing a direct and experimentally testable connection among non-inertial dynamics, nonlinear response, and topological symmetry breaking across platforms ranging from ultracold atoms and exciton-polariton condensates to cosmological analog systems.
\end{abstract}

\maketitle

Description of the interaction between a moving localized perturbation (impurity) and a quantum many-body environment constitutes a canonical problem in condensed matter physics, with profound implications for our understanding of dissipation at the microscopic level~\cite{Dalibard2011, Grusdt_2017}.
The historical foundation of this field was established by Lev Landau~\cite{Landau1941}, who formulated a microscopic criterion for superfluidity, linking the absence of dissipation during impurity motion to the structure of the excitation spectrum. 
According to this criterion, the critical velocity \(v_{c}\) is determined by the minimum ratio of excitation energy to momentum, which for a phonon spectrum in dilute gases corresponds to the speed of sound \(c_s\) \cite{Zwierlein2005, Hadzibabic2006}. 
However, real-world systems, such as liquid helium or ultracold atomic gases, exhibit more intricate scenarios of superfluidity breakdown driven by the nucleation of quantized vortices~\cite{Frisch1992, Saito2001} or radiative deceleration via quasiparticle emission.

A pivotal milestone in the theoretical description of these processes was reached by Lev Pitaevskii~\cite{Pitaevskii1959}, who introduced a method to evaluate the drag force exerted on an impurity moving with the velocity $V$, derived from the density perturbations of the condensate within the framework of the Gross-Pitaevskii equation (GPE)~\cite{Roberts2005}. 
It was demonstrated that in the supersonic regime (\(V>c\)) (with $c$ being the phase velocity of BEC phonon-like excitations), a particle generates a Cherenkov cone of phonons~\cite{Astrakharchik2004}, which serves as the primary channel for energy dissipation. 
This approach was proven to be a versatile tool, forming the basis for the modern theory of Bose polarons and the dynamics of ions in superfluids. 
Yet, despite these advances, the established Landau-Pitaevskii paradigm is fundamentally built upon the assumption of uniform, rectilinear motion. 
A comprehensive framework for quantum friction under non-inertial, accelerated motion has remained a long-standing open challenge~\cite{Schmidt2018, Oppong2019, Sonin2016}.

In recent years, research interest has shifted toward low-dimensional systems, particularly two-dimensional (2D) Bose-Einstein (quasi-)condensates (BECs). 
The 2D geometry imposes fundamental constraints on long-range order, rendering the system highly sensitive to phase fluctuations \cite{Volovik2009, Nakano2024}. 
In 2D condensates, energy dissipation acquires distinctive features: planar phonon emission modifies the spatial structure of shock waves and leads to logarithmic scaling of the effective mass with system size. 
Furthermore, the Berezinskii-Kosterlitz-Thouless transition introduces new energy scales associated with the dynamics of vortex-antivortex pairs \cite{Williams2010}, which can be induced by a moving impurity even at velocities below the formal Landau threshold~\cite{Akram2021}.

The modern experimental frontier has made these questions of non-inertial dynamics experimentally urgent and practically significant.
Contemporary platforms allow unprecedented control over superfluid quantum liquids and superfluidity criteria in various 2D systems. 
For instance, atomic condensates and exciton-polariton condensates have recently become frontier systems for studies of superfluidity. 
Starting from the fundamental works confirming superfluidity at high temperatures by studying friction of moving condensates in the presence of impurity perturbattions~\cite{amo2009superfluidity, lerario2017room} to observation of quantized vorticies~\cite{lagoudakis2008quantized}, quantum half-vortcies~\cite{rubo2007half, lagoudakis2009observation}, vortex clusters~\cite{panico2023onset}, hydrodinamical solitons~\cite{amo2011polariton}, supersolidity~\cite{recati2023supersolidity, trypogeorgos2025emerging, meng2026hybrid, kozhevin2025supersolidity}, combing it with potential practical importance in quantum and optical computing~\cite{kavokin2022polariton, berloff2017realizing}.

This experimental progress has brought a specific, critical problem to the forefront: describing an impurity undergoing a complex, non-rectilinear trajectory has recently acquired particular significance due to advances in optical trapping. 
In contemporary experiments using optical tweezers, impurity atoms often exhibit precessional or circular motion. 
While the rectilinear case is well-studied, composite motion, combining a translational drift at velocity \(V\) with a circular orbital motion of radius \(a\) and frequency \(\omega\), represents a fundamentally different physical regime. 
From the perspective of classical electrodynamics, such a particle is an analog of an accelerated charge emitting synchrotron radiation.
In a quantum fluid, the centripetal acceleration inherent in such motion can render the classical Landau criterion fundamentally inapplicable, thereby opening a new chapter in quantum tribology. 

We demonstrate that the continuous change in the velocity vector allows the impurity to bridge the energy-momentum gap required for phonon excitation. 
Even in the deep subsonic regime (\(V<c\)), a rotating impurity ``pumps'' energy into the medium, emitting phonons at discrete harmonics of the rotational frequency. 
This phenomenon, which we may identify as a quantum analogy of bremsstrahlung, implies that acceleration acts as an effective source of energy and momentum exchange, giving rise to a finite friction coefficient where none would exist in uniform motion.
This effect is expected to be a critical energy-loss channel in high-precision quantum measurements and qubit-impurity platforms.

Another conceptually profound aspect uncovered by our work is the emergence of transverse forces in the absence of classical viscosity.
In classical hydrodynamics, a rotating body in a flow experiences a Magnus force due to the entrainment of the boundary layer.
We show that in a superfluid BEC, an anomalous transverse force arises through a purely quantum mechanical mechanism.
Within the Pitaevskii approach, we contend that this quantum Magnus-like force emerges as a second-order effect in the response theory.
The nonlinearity of the GPE ensures that the accelerated impurity drags the surrounding condensate into motion, generating an effective, self-consistent circulation around the defect site~\cite{Sonin1997}.

This nonlinear response to the accelerated motion forms a dynamically asymmetric density cloud. 
The interaction between the impurity and this induced ``phonon wake'' produces a force component perpendicular to the instantaneous velocity vector. 
Crucially, this force is non-dissipative and performs no work, distinguishing it sharply from its classical counterpart and highlighting its topological origin.
It is highly sensitive to the condensate’s nonlinear coupling constant and the background density, marking it as a collective many-body effect.
\begin{figure}[t!]
\includegraphics[width=0.99\columnwidth]{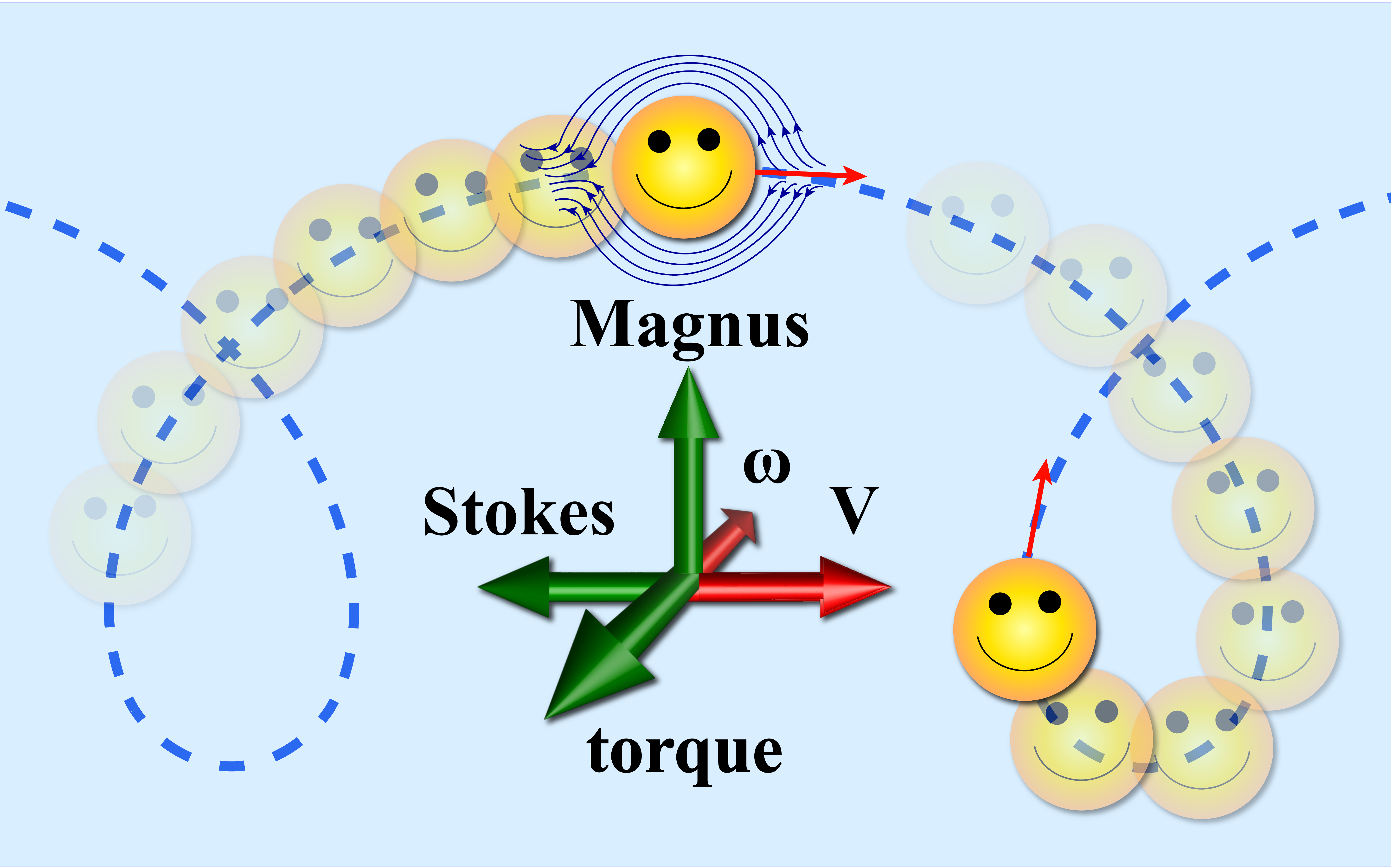} 
\caption{System schematic: an impurity particle undergoes composite translational and orbital motion within a two-dimensional Bose-Einstein condensate. This dynamics gives rise to both longitudinal (Stokes) and transverse (Magnus) forces, alongside a back-action friction torque exerted by the condensate on the moving impurity.}
\label{Fig1}
\end{figure}
%
%
%



We consider a probe impurity particle moving within a 2D BEC (Fig.~\ref{Fig1}). 
The particle moves with a constant translational velocity $\mathbf{V}$ while simultaneously rotating in a circle of radius $a$ with an angular frequency $\omega$. 
The angular velocity vector $\bm{\omega}$ is perpendicular to $\mathbf{V}$, such that the motion occurs entirely within the 2D plane. 
The impurity-BEC interaction is described by the potential $U(\mathbf{r}, t) = U(\mathbf{r} - \mathbf{R}(t))$, where the impurity trajectory $\mathbf{R}(t)$ is defined as $\mathbf{R}(t) = \mathbf{V}t + a(\hat{\mathbf{x}}\cos(\omega t) + \hat{\mathbf{y}}\sin(\omega t))$. 
This non-inertial trajectory introduces a fundamentally new energy scale into the quantum friction problem.

The dissipative effects exerted by the condensate on the impurity are characterized by the drag force $\mathbf{F}$ responsible for friction 
and the torque $\mathbf{T}$ on the moving particle, ${\bf T}||{\bm \omega}$. 
The drag force consists of two contributions: 
The first one is the longitudinal Stokes friction, associated with dissipative processes in the condensate, and directed opposite to the translation velocity of impurity, ${\bf F}_\textrm{S}\propto-{\bf V}$. 
The second contribution is the transverse, anomalous force, a quantum analogue of the classical Magnus effect, assosiated with a non-dissipative processes, and directed perpendicular to impurity longitudinal motion, ${\bf F}_\textrm{M}\propto[{\bf V}\times{\bm\omega}]$. 
The torque arises due to the asymmetry of the induced density distribution relative to the instantaneous center of rotation.
The drag force and torque are defined as~\cite{Astrakharchik2004}:
\begin{gather}
    \label{EqForce01}
    \mathbf{F} = - \left\langle\int \delta n(\mathbf{r}, t) \nabla U(\mathbf{r}, t) d{\bf r}\right\rangle,\\
    \label{EqTorque01}
     \mathbf{T} = - \left\langle\int \delta n(\mathbf{r}, t) [(\mathbf{r} - \mathbf{V}t) \times \nabla U(\mathbf{r}, t)] d{\bf r}\right\rangle,
\end{gather}
where $\langle...\rangle$ stands for the averaging over the rotating period of probe impurity, and $\delta n({\bf r},t)$ is a density response of the media. 

%
%
%
\begin{figure*}[t!]
\includegraphics[width=1.75\columnwidth]{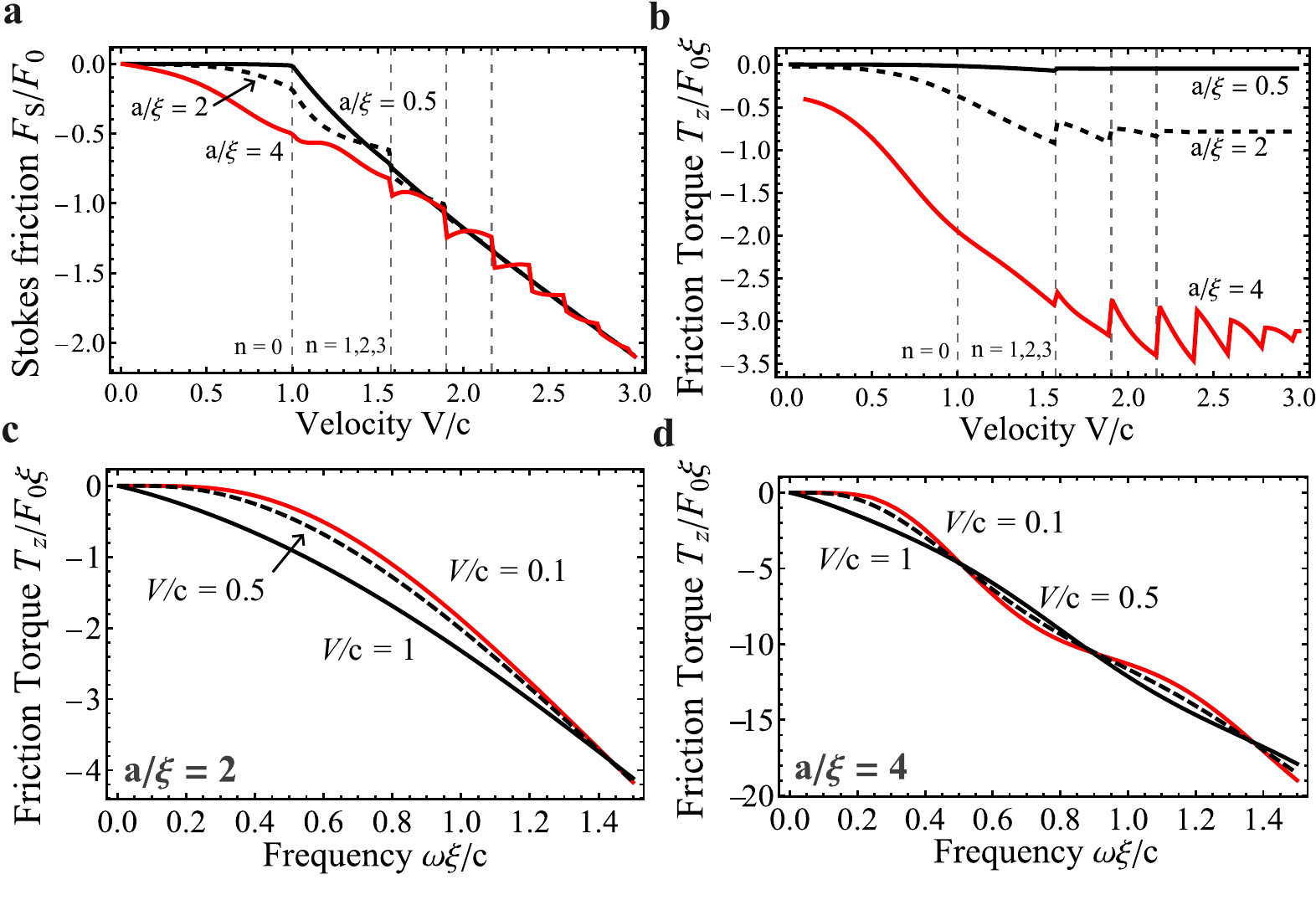} 
\caption{Behavior of the Stokes friction force and torque: generalization of the Landau criterion to non-inertial motion. 
(a) Normalized friction force $F_{\rm S}/F_0$ and (b) torque $T_z/F_0\xi$ as functions of the dimensionless impurity velocity $V/c$ for different ratios of the rotation amplitude to the healing length: $a/\xi = 4$ (red solid), $0.5$ (black dashed), and $2$ (black solid). 
The latter corresponds to $\omega a \ll c$, where the rotation is small, and the  Landau criterion $V > c$ for the friction force is largely satisfied. 
Vertical dashed lines indicate the critical resonant velocity thresholds corresponding to the angular harmonics $n = 0, 1, 2, 3$. 
The corresponding critical velocities are $V_c = c$, $V_c \approx 1.58c$, $V_c \approx 1.9c$, and $V_c \approx 2.16c$. The curves are evaluated for the fixed rotation frequncy $\omega \xi/c = 0.25$.
Panels (c) and (d) show torque as a function of the normalized frequency $\omega\xi/c$ for fixed parameter $a/\xi = 2$ (c) and $a/\xi = 4$ (d). 
Different curves correspond to $V/c = 0.1$ (red solid), $0.5$ (black dashed), and $1.0$ (black solid); $F_0 = n_c U_0^2 / m c^2 \xi^3$.}
\label{Fig2}
\end{figure*}
%
%
%


\textit{\textcolor{blue}{First-order density response: longitudinal drag force and torque.---}}The dynamics of the condensate is governed by the Gross-Pitaevskii equation. 
From the linear response theory, the Fourier component of the BEC density correction $\delta n^{(1)}(k)$ (in what follows, we will use the notation $k\equiv({\bf k},\Omega)$ for Fourier harmonics) is related to the moving impurity potential through the linear response function $\chi^{(1)}(k)$:
\begin{equation}
\delta n^{(1)}(k) = \chi^{(1)}(k) U(k).
\end{equation}
Applying the Jacobi-Anger expansion to the potential in $\bf k$-space yields the following Fourier representation:
\begin{equation}
U(k) = 2\pi U_{\bf k} \sum_{n=-\infty}^{\infty} (-i)^n J_n(ka) e^{in\phi_k}\delta(\Omega-\omega_{n{\bf k}}),
\end{equation}
where $\phi_k$ is the polar angle of the wave vector $\mathbf{k}$, $J_n(ka)$ are the Bessel functions, and $\omega_{n{\bf k}}=\mathbf{k}\cdot\mathbf{V} + n\omega$.
The susceptibility of the BEC at zero temperature can be extracted from the linearized GPE (see Supplemental Material~\cite{SMBG}):
\begin{equation}
\label{EqSuccept00}
\chi^{(1)}(k) = \frac{n_c \textbf{k}^2 / m}{(\Omega + i0)^2 - \epsilon_\textbf{k}^2},
\end{equation}
where $\epsilon_\textbf{k} = \sqrt{\frac{k^2}{2m}(\frac{k^2}{2m} + 2gn_c)}=ck\sqrt{1+k^2\xi^2}$ is the Bogoliubov excitation spectrum with $c^2=gn_c/m$ the sound velocity in BEC, and $\xi=1/2mc$ the healing length; $n_c$ is the equilibrium density of the condensate, $g$ is the interacting potential of Bose particles with mass $m$.

Substituting Eq.~\eqref{EqSuccept00} in Eqs.~\eqref{EqForce01} and~\eqref{EqTorque01}, and then averaging over the period of impurity rotation, the longitudinal drag (Stokes) force and the torque acquire the form: 
\begin{gather}
\label{generalFM01}
\mathbf{F}_\textrm{S} = \sum_{n=-\infty}^{\infty}\int \frac{d{\bf k}}{(2\pi)^2}  \mathbf{k}  |U_{\bf k}|^2 J_n^2(ka)\, \text{Im} \chi^{(1)}({\bf k}, \omega_{n{\bf k}}),
\\
\label{generalFM02}
\mathbf{{T}} = \hat{\mathbf{z}} \sum_{n=-\infty}^{\infty} n\int \frac{d{\bf k}}{(2\pi)^2} |U_{\bf k}|^2 J_n^2(ka) \text{Im} \chi^{(1)}({\bf k}, \omega_{n{\bf k}}).
\end{gather}
These expressions are of general form and can be applied to any medium whose response is characterized by the linear susceptibility $\chi^{(1)}(k)$. 
In the case of a weakly interacting BEC, the imaginary part of the susceptibility,
\begin{eqnarray}
\label{EqSucceptIm}
\text{Im}\,\chi^{(1)}(k)
=-\pi \frac{n_c\textbf{k}^2/m}{2\epsilon_\textbf{k}}
\left[\delta(\Omega-\epsilon_\textbf{k})-\delta(\Omega+\epsilon_\textbf{k})\right], 
\end{eqnarray}
determines the conditions for emission of elementary excitations (in perturbed BEC). 
Each term $n$ in the summation corresponds to the emission or absorption of $n$ quanta of rotation. 
The condition for non-zero friction is governed by the resonance condition: $\omega_{n{\bf k}}\equiv \mathbf{k} \cdot \mathbf{V} + n\omega = \pm \epsilon_\textbf{k}$. 

In the absence of rotation ($\omega=0$) and translational motion ($V=0$), and in the case of contact interaction $U_{\bf k}=U_0$, the force and the torque read:
\begin{gather}
\label{limCases01}
(F_\textrm{S})_x = -n_cm^{{2}}U_0^2\frac{V^2-c^2}{V}\,\theta[V-c],\\
\label{limCases02}
T_z=-\text{sgn}(\omega)mn_cU_0^2
\sum_{n=1}^{\infty}n\left(1-\frac{1}{\alpha_n}\right)J^2_n(k_na),
\end{gather}
where $\alpha_n=\sqrt{1+n^2(\omega/mc^2)^2}$ and $k_n$ is determined by the resonance condition $\epsilon_\textbf{k}=n\omega$, yielding
$k_n=\Bigl(\sqrt{\sqrt{1+(n\omega/mc^2)^2}-1}\Bigr)/(\xi\sqrt{2})$. 
Evidently, the force~\eqref{limCases01} emerges due to the generation of excitations in the BEC by the moving impurity only when 
$V>c$, satisfying the Landau criterion, i.e., expressing the condition that the drag force appears due to the creation of BEC excitations from Cherenkov radiation, as indicated by the presence of the imaginary part of the response function in Eqs.~\eqref{generalFM01} and~\eqref{generalFM02}. 
In this regime, the force coincides with that found in~\cite{Astrakharchik2004}. 
In contrast, the rotational friction~\eqref{limCases02} exists at any value of tangential velocity, $a\omega$, and it is thresholdless. 

The conventional Landau criterion $V>c$ for the translational impurity velocity ${\bf V}$ becomes inapplicable in the presence of impurity rotation $\omega\neq0$, as follows from Eqs.~\eqref{generalFM01} and~\eqref{generalFM02}. 
Acceleration fundamentally modifies the kinematic constraints on quasiparticle emission.
To demonstrate this unique difference analytically at $V<c$, consider the limit of small radius $a\rightarrow 0$ ($ka\ll1$).
Then, only $n=\pm1$ harmonics mainly contribute to the sum in~\eqref{generalFM01} and ~\eqref{generalFM02}, and taking linear dispersion of the Bogoliubov quasi-particles $\epsilon_\textbf{k}\approx ck$, the drag force and torque read (see the derivation in~\cite{SMBG}):
\begin{gather}
(F_\textrm{S})_x = -\frac{3}{64}\left(\frac{U_0^2n_ca^2}{mc^2}\right)
\frac{V|\omega|^5}{c^6}\frac{8+12\frac{V^2}{c^2}+\frac{V^4}{c^4}}{\left(1-\frac{V^2}{c^2}\right)^{11/2}},
\\
T_z = -\frac{\textrm{sgn}(\omega)}{64}\left(\frac{U_0^2n_ca^2}{mc^2}\right)\left(\frac{\omega}{c}\right)^4\frac{8+24\frac{V^2}{c^2}+3\frac{V^4}{c^4}}{\left(1-\frac{V^2}{c^2}\right)^{9/2}}.
\end{gather}
In the limit of small translational velocity, $V\rightarrow0$, the force vanishes as $F_\textrm{S}\propto V$, but the torque coincides with Eq.~\eqref{limCases02} as it originates from $n=\pm1$ harmonics given a small argument of the Bessel function, $k_{\pm1}a\ll1$. 



A general analysis for arbitrary parameters requires a numerical treatment. 
Assuming, without loss of generality, that the translational velocity is along the $x$-axis, ${\bf V}=(V,0)$, and using Eq.~\eqref{EqSucceptIm}, we can find the force and the torque~\cite{[{See Supplemental Materials at [URL], which gives the details of the derivations}]SMBG}.
Figure~\ref{Fig2} (a) and (b) show the dependence of the Stokes friction force and torque on the velocity ${\bf V}$ and frequency $\omega$. 
It serves as the demonstration of: (i) the generalization of the Landau criterion for the friction force at finite frequencies, as evidenced by the onset of a non-zero force in the subcritical regime $V<c$ when $\omega\neq0$ -- a regime where the conventional paradigm predicts strictly zero dissipation; and (ii) the emergence of a quantum stick-slip behaviour in both the force and torque. 
These specific features can be understood qualitatively. 
Performing the angular integration in Eq.~\eqref{generalFM01} and~\eqref{generalFM02} yields a series of threshold conditions for the excitations:
\begin{gather}\label{conditions}
(kV)^2\geqslant(\epsilon_\textbf{k}+n\omega)^2,
\end{gather}
where $n$ is the rotational motion harmonics. 
For $n=0$, this expression recovers the  Landau criterion. 
At finite $\omega$, negative values of $n$ extend the kinematic threshold, rendering it possible to excite quasiparticles and thereby enabling finite friction even in the subcritical regime, $V<c$. 
Conversely, positive values of $n$ yield a discrete set of effective threshold conditions, each corresponding to a sharp jump in both the friction force and torque.

A compelling analogy can be drawn from condensed matter physics: the index $n>0$ can be interpreted as labeling a set of quantum well subbands (Landau levels) with effective energies $n\omega+\epsilon_k$. 
This effectively partitions the Bose gas into a multicomponent subsystem, in which each subband has its own Landau-like criterion. 
Consequently, the total contribution to the friction force mimics the Landauer formula for quantum-well conductivity, expressed as a sum over channels. 
The onset of dissipation in each individual 'subband' is thus manifested as a distinct step-like discontinuity in the friction force and torque.

This effect can be characterized by two dimensionless parameters: $a/\xi$ (the ratio of the radius of rotation to the healing length) and $\omega\xi/c$ (the dimensionless angular frequency). 
The latter primarily dictates the energy spacing between the subbands. Meanwhile, their product (the effective rotation tangential velocity $\omega a/c$) determines the magnitude of the jumps. As shown by the black solid lines in Figs.~\ref{Fig2}(a) and \ref{Fig2}(b), in the limiting case where $\omega a/c \ll 1$, the sharp jumps disappear; the behavior of the Stokes friction is then nearly described by Eq.~\eqref{limCases01}, and the friction torque vanishes (see also other examples of such behavior for various rotation parameters in~\cite{SMBG}).

Furthermore, we investigated the dependence of the friction torque on the rotation frequency at fixed translational velocities, $V/c$. 
Figures~\ref{Fig2}(c) and \ref{Fig2}(d) summarize these results for two effective rotation amplitudes, $a/\xi$, across multiple values of $V/c$. 
When $a/\xi = 2$, the friction torque increases monotonically with rotation frequency, as expected. However, for rotations with a larger amplitude ($a/\xi = 4$), the torque begins to exhibit shallow oscillations. 
These oscillations become particularly pronounced in the low-velocity limit, $V \ll c$ (Fig.~\ref{Fig2}(d)).



\begin{figure}[tb!]
\includegraphics[width=0.99\columnwidth]{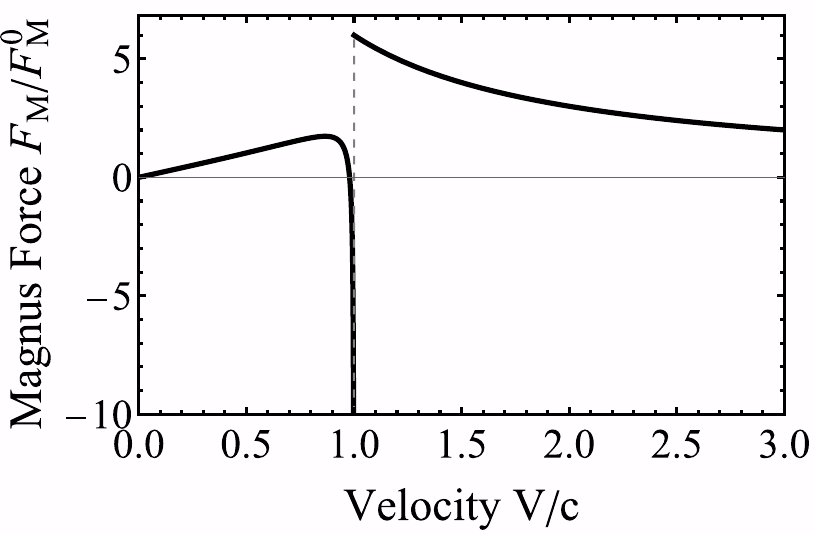} 
\caption{Normalized anomalous transverse force $F_{\rm M}/F_{\rm M}^0$ as a function of dimensionless translational velocity $V/c$. 
Here, $F_{\rm M}^0 =  U_0^3\pi a^2k_0^{10}n_c/(96  m^2c\omega^3)$. }
\label{Fig3}
\end{figure}

\textit{\textcolor{blue}{Second-order response: Anomalous transverse quantum Magnus force.---}}The transverse drag force is absent in the first-order density response. 
Indeed, in the first order with respect to the BEC-impurity potential, the perturbed density has axial symmetry with respect to the direction along ${\bf V}$. 
This symmetry, however, is dynamically broken when the impurity drags the BEC density along a circular trajectory, generating an effective self-consistent circulation.
This effect appears in the second order:
\begin{gather}
{\bf F}_\textrm{M}
=i\sum_{\bf k}{\bf k}_{\perp}\left\langle\delta n^{(2)}({\bf k},t)U(-{\bf k},t)\right\rangle\\\nonumber
=-i\sum_k{\bf k}_{\perp}U^*(k)\delta n^{(2)}(k),
\end{gather}
where ${\bf k}_\perp$ indicates the component perpendicular to the translation velocity ${\bf V}=(V_x,0)$, thus ${\bf k}_\perp=(0,k_y)$. 
Here,
\begin{eqnarray}
\delta {n}^{(2)}(k)=(2\pi)^2\sum_{k_1,k_2}\delta_{k,k_1+k_2}\chi^{(2)}(k_1,k_2)U(k_1)U(k_2)~~
\end{eqnarray}
being the Fourier transforms of
the second-order density response. 
Real-valuedness of external potential and the density dictates $U^*(k)=U(-k),\,\delta n^{(2)}(k)^*=\delta n^{(2)}(-k)$, and $\chi^{(2)}(k_1,k_2)=\chi^{(2)}(k_2,k_1),\,\chi^{(2)*}(k_1,k_2)=\chi^{(2)}(-k_1,-k_2)$. 
Then, the Magnus force can be written in the form: 
\begin{eqnarray}
{\bf F}_\textrm{M}=-i\sum_{k_1,k_2}({\bf k}_{12})_{\perp}X(k_1,k_2)\chi^{(2)}(k_1,k_2),
\end{eqnarray}
where
\begin{eqnarray}
&&X(k_1,k_2)=U^*(k_{12})U(k_1)U(k_2)\\
\nonumber 
&&~~~=-U_0^3\sum_{n_1, n_2} J_{n_1+n_2}(k_{12} a) J_{n_1}(k_1 a) J_{n_2}(k_2 a)\\
\nonumber
&&~~~~~\times
(2\pi)^3
e^{in_1(\phi_1-\phi_{12})}e^{in_2(\phi_2-\phi_{12})}\\
\nonumber
&&~~~~~\times\delta(\Omega_1 - \mathbf{k}_1\mathbf{V} - n_1\omega)\delta(\Omega_2 - \mathbf{k}_2\mathbf{V} - n_2\omega),
\end{eqnarray}
with $k_{12}=({\bf k}_1+{\bf k}_2,\Omega_1+\Omega_2)$ and assuming the contact interacting potential, $U_{\bf k}=U_0$. 
In this expression, all harmonics are of equal importance. However, an analytical result can be obtained in the limit of a small rotation radius (\(ka \ll 1\)) and high-frequency rotation (\(\omega \gg ck, Vk\)) for an arbitrary relation between \(V\) and \(c\), analogous to the Stokes drag force. 
After derivations, we find~\cite{SMBG}:
\begin{eqnarray}
(F_\textrm{M})_y &=&{\pi n_ca^2k_0^3}\frac{(U_0k_0)^3}{2^4(mc^2)^2}\frac{ck_0}{V}\left(\frac{ck_0}{\omega}\right)^3 \\
\nonumber
&&\times
\left[1 - \frac{c^2 - \frac{5}{6}V^2}{c\sqrt{c^2 - V^2}}\theta(c-V) \right], 
\end{eqnarray}
where the $k$-integration was cut by the lowest values of $k_0\sim1/\xi$. 
This expression demonstrates linear growth of the anomalous transverse force at low translational velocities (\(V \ll c\)) and exhibits a specific jump at \(V/c = 1\), as shown in Fig.~\ref{Fig3}.

%
%

%
%
%

It is important to emphasize that, unlike the dissipative friction force, this anomalous transverse force is not governed by dissipative processes~\cite{SMBG}. 
Moreover, since the Magnus force acts perpendicular to the translational velocity $\mathbf{V}$, it performs no work.  
This non-dissipative, geometry-induced character distinguishes it fundamentally from the classical Magnus effect and places it within the broader class of anomalous Hall-like responses in quantum fluids. {Physically, the discussed quantum Magnus force arises from the friction between a rotating particle and the Bose fluid. The similar force appears due to the radiative friction between a rotating charged particle and an electromagnetic field~\cite{PhysRevLett.134.175001}.}



\textcolor{blue}{\textit{Conclusions.---}}We investigated the non-inertial dynamics of a massive external object, specifically an impurity, immersed in a quantum fluid modeled via the Bogoliubov framework for a weakly interacting Bose gas. 
By employing a combination of analytical methods and numerical simulations, we demonstrated that the resulting friction force fundamentally transcends the conventional Landau criterion, which strictly holds only for uniform, unaccelerated inertial motion. Our work establishes a generalized framework for quantum friction that naturally recovers the classical Landau-Pitaevskii paradigm in the limit of vanishing acceleration.
Furthermore, we show that both the calculated friction force and the induced torque exhibit distinct quantum stick-slip transitions.
This phenomenon indicates the emergence of a discrete family of Landau-like critical conditions for each harmonic of the impurity's rotational motion.
Finally, we find that the simultaneous breaking of time-reversal and space-reversal symmetries gives rise to a novel, anomalous transverse quantum force:
there emerges a non-dissipative Magnus-like hydrodynamic response of the quantum fluid, which acts directly on the non-inertially moving impurity, revealing a deep connection between topological symmetry breaking and dissipation in driven quantum systems. 



\textcolor{blue}{\textit{A universal platform for quantum tribology.---}}
The state-of-the-art experimental platforms have reached the level of sophistication required to observe these subtle radiative and transverse effects. 
By employing deterministic state preparation of single \({}^{87}\)Rb atoms in a 2D \({}^{41}\)K condensate, researchers can now utilize Bragg spectroscopy and in-situ imaging to map the density perturbations with sub-micron resolution. 
Such experiments can directly measure the angular distribution of the `bremsstrahlung' phonons and the resulting displacement of the impurity caused by the Magnus force.

However, the implications of our results extend significantly beyond the realm of ultracold atomic gases. 
Our framework is universal and applies to any nonlinear quantum fluid with a well-defined quasiparticle spectrum.
In particular, the dynamics described here is directly applicable to exciton-polariton condensates in semiconductor microcavities~\cite{Carusotto2006}, where the driven-dissipative nature of the system may further enhance the predicted patterns. 
Moreover, our study provides a potential laboratory analog for cosmological processes. 
In models where dark matter consists of ultralight bosons forming a galactic-scale BEC, the movement of baryonic structures (stars or black holes) through this background would be subject to the same drag and transverse forces evaluated here. 
The predicted dependence of the anomalous transverse force on the medium's nonlinearity could thus provide a new theoretical tool for exploring the interaction mechanisms between dark and baryonic matter in the early universe.


\textit{Acknowledgments.}
This work was supported by the National Natural Science Foundation of China (NSFC) under Grant No.~W2532001, the Guangdong Basic and Applied Basic Research Foundation under Grant No.~2026A1515012415, the Ministry of Science and Higher Education of the Russian Federation (Project FSUN-2026-0004), and the Foundation for the Advancement of Theoretical Physics and Mathematics ``BASIS''.
We are grateful to Elizaveta Osipova for help with the figures.

\bibliography{biblio}
\bibliographystyle{apsrev4-2}


\end{document}


\title{Supplementary Material for ``Quantum tribology: acceleration-induced Stokes friction and Magnus force in correlated Bose fluids''}





\author{V.~M.~Kovalev}
\affiliation{Guangdong Technion -- Israel Institute of Technology, 241 Daxue Road, Shantou, Guangdong, China, 515063}
\affiliation{Rzhanov Institute of Semiconductor Physics, Siberian Branch\\ of Russian Academy of Science, Novosibirsk 630090, Russia}
\affiliation{Novosibirsk State Technical University, Novosibirsk 630073, Russia}

\author{A.~N.~Osipov}
\affiliation{Technion -- Israel Institute of Technology, Haifa, 3200003, Israel}
\affiliation{Guangdong Technion -- Israel Institute of Technology, 241 Daxue Road, Shantou, Guangdong, China, 515063}


\author{I.~G.~Savenko}
\affiliation{Guangdong Technion -- Israel Institute of Technology, 241 Daxue Road, Shantou, Guangdong, China, 515063}
\affiliation{Technion -- Israel Institute of Technology, Haifa, 3200003, Israel}

\date{\today}

\begin{abstract}
This Supplementary Material provides a comprehensive derivation of the nonlinear response functions for a correlated Bose fluid governed by the Gross-Pitaevskii equation. 
Employing the Madelung transformation together with a second-order expansion in the external potential, we obtain a closed-form expression for the quadratic susceptibility. 
Building on this framework, we analyze the Stokes friction force and torque experienced by a rotating particle and demonstrate that the corresponding resonance conditions naturally extend the Landau criterion. 
Remarkably, contributions from negative angular harmonics give rise to a finite friction force even below the Landau threshold. 
In the long-wavelength regime, we derive an explicit formula for the non-dissipative anomalous transverse Magnus-like force (which we will call Magnus force for brevity), which disappears when both rotation and translation are absent. 
Finally, in the limit of rapid rotation, the Magnus force collapses to a compact analytical form and displays a linear dependence on velocity at low speeds, revealing an effective transverse viscous drag.
\end{abstract}

\maketitle

\tableofcontents


\section{Derivartion of the second-order response functions}


The Gross-Pitaevskii equation describing the condensate and accounting for an external perturbing potential $U(\mathbf{r}, t)$ reads as~\cite{gross1961structure, pitaevskii1961vortex}:
\begin{eqnarray}
i\hbar\frac{\partial\psi}{\partial t}=\left(-\frac{\hbar^2}{2m}\nabla^2+g|\psi|^2+U(\mathbf{r},t)\right)\psi,    
\end{eqnarray}
where $m$ is the boson mass and $g$ is the interaction constant.
Applying the Madelung transformation $\psi(\mathbf{r},t)=\sqrt{n(\mathbf{r},t)}e^{i\theta(\mathbf{r},t)}$ and introducing the velocity field $\mathbf{v}=(\hbar/m)\nabla\theta$ (disregarding the quantum pressure in the long-wavelength limit $k\xi \ll 1$), we obtain the system of equations, consisting of the continuity equation,
\begin{eqnarray}
\nonumber
\frac{\partial n}{\partial t}+\nabla\cdot(n\mathbf{v})=0,   
\end{eqnarray}
and the Euler equation,
\begin{eqnarray}
\nonumber
m\frac{\partial\mathbf{v}}{\partial t}+\nabla\left(\frac{mv^2}{2}+gn+U(\mathbf{r},t)\right)=0.   
\end{eqnarray}

Furthermore, we can expand the variables in orders of smallness of the external potential,
$$n=n_c+n_1+n_2+\dots,$$
$$\mathbf{v}=\mathbf{v}_1+\mathbf{v}_2+\dots,$$
where $n_c$ is the equilibrium uniform density, and $\mathbf{v}_0=0$. 
Next, we transform to Fourier space with the 4-vector $k=(\Omega,\mathbf{k})$.


Keeping the first-order terms in $U$, we obtain:
$$-i\Omega n_1+in_0\mathbf{k}\cdot\mathbf{v}_1=0,$$
$$-i\Omega m\mathbf{v}_1+i\mathbf{k}(gn_1+U)=0.$$
From the first equation, we can express the velocity:
$$\mathbf{v}_1(k)=\frac{\omega\mathbf{k}}{n_ck^2}n_1(k).$$
Then, substituting this in the second equation, we find the retarded linear susceptibility,
\begin{eqnarray}
\chi^{(1)}(k)=\frac{n_c\textbf{k}^2/m}{(\Omega+i\delta)^2-c^2k^2},   
\end{eqnarray}
which is defined as $n_1(k)=\chi^{(1)}(k)U(k)$ with $c=\sqrt{gn_0/m}$ the speed of sound in the condensate.


Furthermore, collecting second-order terms (proportional to $U^2$) yields:
$$-i\Omega n_2+in_0\mathbf{k}\cdot\mathbf{v}_2+i\mathbf{k}\cdot\mathcal{F}[n_1\mathbf{v}_1]=0,$$
$$-i\Omega m\mathbf{v}_2+i\mathbf{k}\left(gn_2+\frac{m}{2}\mathcal{F}[v_1^2]\right)=0,$$
where $\mathcal{F}$ denotes the Fourier transform.
We can express $\mathbf{k}\cdot\mathbf{v}_2$ from the second-order Euler equation and substitute it in the continuity equation, multiplying by $\Omega$:
$$(\Omega^2-c^2k^2)n_2=n_0\frac{k^2}{2}\mathcal{F}[v_1^2]+\Omega\mathbf{k}\cdot\mathcal{F}[n_1\mathbf{v}_1].$$

Next, we calculate the Fourier transforms of the quadratic terms:
$$\mathcal{F}[v_1^2]_k=\int\frac{\Omega_1\Omega_2(\mathbf{k}_1\cdot\mathbf{k}_2)}{n_c^2k_1^2k_2^2}n_1(k_1)n_1(k_2),$$
$$\Omega\mathbf{k}\cdot\mathcal{F}[n_1\mathbf{v}_1]=\frac{\Omega}{2n_c}\int\left(\frac{\Omega_1(\mathbf{k}\cdot\mathbf{k}_1)}{k_1^2}+\frac{\Omega_2(\mathbf{k}\cdot\mathbf{k}_2)}{k_2^2}\right)n_1(k_1)n_1(k_2),$$
where the integration is performed over $k_1$ and $k_2$ under the condition $k=k_1+k_2$.
Using $n_1(k_i)=\chi^{(1)}(k_i)U(k_i)$ and $(\Omega^2-c^2k^2)^{-1}=m\chi^{(1)}(k)/(n_0k^2)$, we rewrite $n_2(k)$ in the form $n_2(k)=\int\chi^{(2)}(k;k_1,k_2)U(k_1)U(k_2)$.
Substituting the explicit form of $\chi^{(1)}(k)$ and canceling out common factors, we obtain the explicit expression for the quadratic susceptibility:
\begin{eqnarray}
\label{EqSuccept01}
\chi^{(2)}(k_1,k_2)=
\frac{n_c}{2m^2}
\frac{k^2\Omega_1\Omega_2(\mathbf{k}_1\cdot\mathbf{k}_2)+\Omega\Omega_1k_2^2(\mathbf{k}\cdot\mathbf{k}_1)+\Omega\Omega_2k_1^2(\mathbf{k}\cdot\mathbf{k}_2)}{((\Omega+i\delta)^2-c^2k^2)((\Omega_1+i\delta)^2-c^2k_1^2)((\Omega_2+i\delta)^2-c^2k_2^2)},
\end{eqnarray}
where $\Omega=\omega_1+\omega_2$ and $\mathbf{k}=\mathbf{k}_1+\mathbf{k}_2$.

\newpage
\section{Stokes friction force and torque in the limit $ka\ll1$ and $V<c$}

Here, we analytically demonstrate the extension of the Landau criterion for a small rotation radius, \(ka\ll1\), in the subsonic regime (\(V<c\)). The condition \(ka\ll1\) justifies truncating the harmonic expansion to only \(n=0\) and \(n=\pm1\) terms. The zero harmonic (\(n=0\)) directly corresponds to the Pitaevskii regime with \(\omega=0\), which exerts no net force when \(V<c\). Consequently, the remaining \(n=\pm1\) harmonics yield:%
\begin{gather}
\label{SMweak1}
(F_\textrm{S})_x=-\frac{\pi n_ca^2U_0^2}{4mc}\int\frac{d{\bf k}}{(2\pi)^2}k^4\cos(\phi_k)
\left[\delta(kV\cos(\phi_k)+\vert{}\omega\vert{}-ck)-\delta(kV\cos(\phi_k)+\vert{}\omega\vert{}+ck)\right]\\
\nonumber
=-\frac{\pi n_ca^2U_0^2}{8\pi ^2mcV}
\int k^4dk
\left[\frac{ck-|\omega|}{\sqrt{(kV)^2-(ck-|\omega|)^2}}+
\frac{ck+|\omega|}{\sqrt{(kV)^2-(ck+|\omega|)^2}}\right],
\end{gather}
where the integration limits are determined by the requirement that the radicands remain non-negative. Analyzing these constraints enables direct integration over \(k\), which results in:%
\begin{gather}
\label{SMweak2}
(F_\textrm{S})_x = -\frac{3}{64}\left(\frac{U_0^2n_ca^2}{mc^2}\right)
\frac{V|\omega|^5}{c^6}\frac{8+12\frac{V^2}{c^2}+\frac{V^4}{c^4}}{\left(1-\frac{V^2}{c^2}\right)^{11/2}}.
\end{gather}
A similar analysis yields the corresponding expression for the torque:
\begin{gather}
\label{SMweak3}
T_z = -\frac{\textrm{sgn}(\omega)}{64}\left(\frac{U_0^2n_ca^2}{mc^2}\right)\left(\frac{\omega}{c}\right)^4\frac{8+24\frac{V^2}{c^2}+3\frac{V^4}{c^4}}{\left(1-\frac{V^2}{c^2}\right)^{9/2}},
\end{gather}
where $\textrm{sgn}(\omega)$ is a signum function.

\section{Discussion of general formulas for friction force and torque}

Assuming, without loss of generality, that the velocity is aligned along the $x$-direction: $\mathbf{V} = (V,0)$, and completing the angular integration in Eqs.~(6) and~(7) of the main text, yields the following general expressions for the Stokes friction and  torque:
\begin{eqnarray}
    F_\textrm{S} &=& - \frac{n_c|U_0|^2}{mc^2\xi^3}\int\limits_0^{\infty} \frac{d\tilde k\, \tilde k^4}{\tilde\epsilon_k} 
    \sum_{n=-\infty}^{+\infty} J_n^2(\tilde k a/\xi) (\tilde\epsilon_k+n\tilde\omega)
    \frac{
    \Theta\left(\frac{\tilde k V}{c}-|\tilde\epsilon_k+n\tilde\omega|\right)
    }{\frac{\tilde k V}{c}\sqrt{\left(\frac{\tilde kV}{c}\right)^2 - (\tilde\epsilon_k+n\tilde\omega)^2}},
    \label{F_general} \\
    T_z &=& -\frac{n_c|U_0|^2}{m c^2 \xi^2}\frac{a}{\xi} \int\limits_{0}^{\infty}\frac{d\tilde k\tilde k^3}{\tilde \epsilon_k} \sum_{n=-\infty}^{\infty} n J_n^2(\tilde ka/\xi)
 \frac{\Theta\left(\frac{\tilde k V}{c}-|\tilde\epsilon_k+n\tilde\omega|\right)}{\sqrt{\left(\frac{\tilde kV}{c}\right)^2 - \left(\tilde \epsilon_k + n\tilde\omega\right)^2}}.
 \label{Mz_general} 
\end{eqnarray}
Here, we introduced the dimensionless parameters $\tilde k = k\xi$, $\tilde \omega = \omega \xi/c$, and $\tilde\epsilon_k = \tilde k\sqrt{1+\tilde k^2}$. 
Thus, the friction force and torque, given by Eqs.~\eqref{F_general} and~\eqref{Mz_general}, are expressed as a series of terms. 
Each term takes the form of a product of an $n$-th order Bessel function and a function representing the resonant activation of that specific term at $\tilde k V/c = |\tilde\epsilon_k+n\tilde \omega|$. 
The convergence of the series is governed by the exponential decay of the Bessel functions $J_n(\tilde k)$ for $n \gg \tilde k$, with a natural wave vector cutoff provided by the healing length, $\tilde k_{\rm max} \sim \xi/a$.
The activation condition for each term can be solved in the $(V/c, \tilde k)$ domain, yielding:
\begin{equation}
\label{activation condition}
    \frac{V(\tilde k)}{c} = \sqrt{1+\tilde k^2} +\frac{n\tilde\omega}{\tilde k}.
\end{equation}

\begin{figure*}[!h]
\includegraphics[width=0.5\columnwidth]{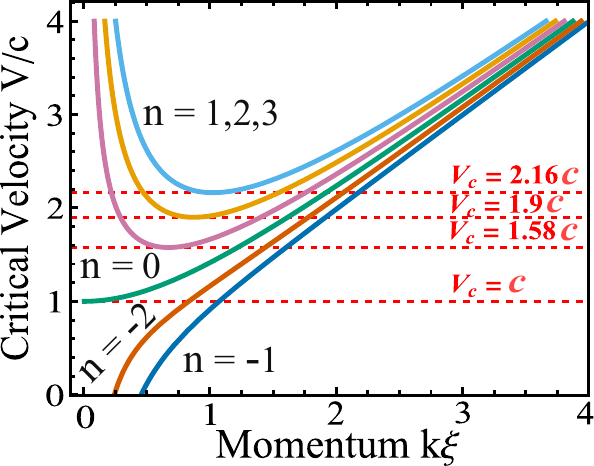} 
\caption{Normalized critical resonant velocity $V/c$ as a function of momentum $k$ for various angular harmonics $n$ given by Eq.~\eqref{activation condition}. 
Solid curves represent the kinematic resonance conditions for the excitation of specific modes ($n = -2, -1, 0, 1, 2, 3$). 
Horizontal red dashed lines indicate the absolute minimum velocity thresholds $V_c$ required to trigger the $n \ge 0$ harmonics. 
These resonance thresholds correspond to $V_c = c$, $V_c \approx 1.58c$, $V_c \approx 1.9c$, and $V_c \approx 2.16c$. 
The curves are evaluated for the fixed parameters $a/\xi = 2$ and $\omega \xi/c = 0.25$, same as for Fig.~2 from the main text.}
\label{FigS1}
\end{figure*}

Figure~\ref{FigS1} illustrates the activation condition~\eqref{activation condition} for the first several contributions ($-2 \leq n \leq 3$) with $\tilde \omega = 0.25$ and $a/\xi = 2$. 
For $n \geq 0$, the corresponding contributions to the friction force and torque are strictly zero when $V < V_c$, where $V_c$ represents the critical velocity thresholds determined by $dV(\tilde k)/d\tilde k = 0$. 
It should be noted, that this equation can be solved analytically.
However, the solutions are cumbersome and they do not provide additional physical insight. 

For $n=0$, the threshold simply recovers the Landau criterion: $V/c > 1$. 
The terms with negative $n$ provide a nontrivial contribution to friction even below the Landau threshold ($V/c < 1$). 
As it is discussed in the main text, the conditions $V>V_c^n$ for $n\geq 0$ could be interpretated as a generalization of the Landau condition for the rotating particle, corresponding to $n_{th}$ rotation subband leading to appeareance of a sharp jumps in friction force and torque velocity dependence. 
Conversely, the thresholdless terms ($n < 0$) correspond to the surpassing, or ``breaking'' of the Landau criterion ($V > c$) due to the rotational component of the motion.
 
\begin{figure}[!h]
\includegraphics[width=0.95\columnwidth]{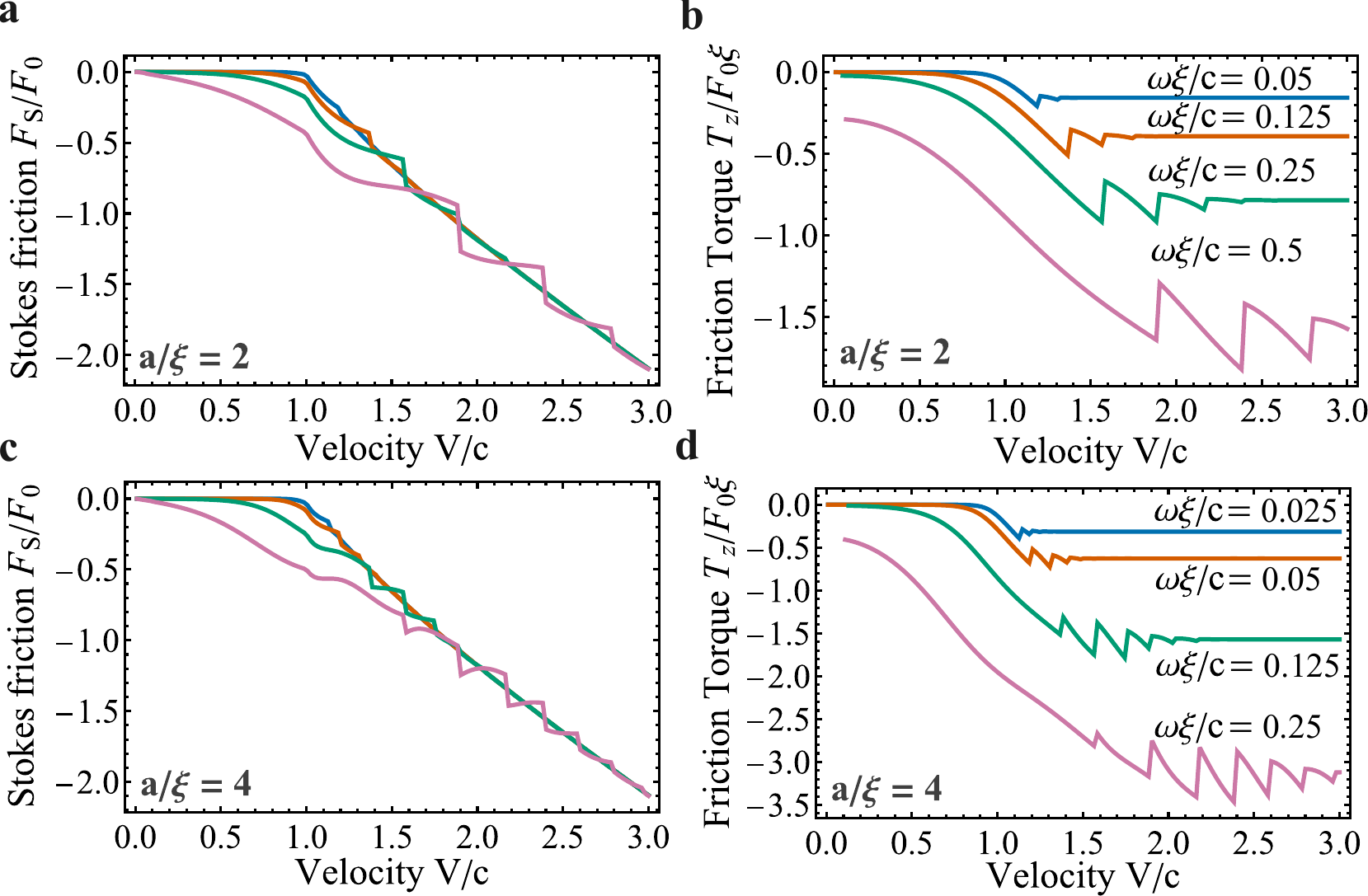} 
\caption{(a, c) Normalized Stoke's friction force $F_{\rm S}/F_0$, and (b, d) torque $T_z/F_0\xi$ as functions of the dimensionless velocity $V/c$. 
The panels are grouped by the ratio of the healing length and the rotation amplitude, with the upper panels (a, b) evaluated at $a/\xi = 2$ and the lower panels (c, d) evaluated at $a/\xi = 4$. 
The solid curves correspond to different normalized rotation frequencies $\omega\xi/c$, which are indicated in the figure. 
The force normalization constant is $F_0 = n_c U_0^2 / m c^2 \xi^3$.}
\label{FigS2}
\end{figure}

Figure~\ref{FigS2} demonstrates the dependence of the Stoke's friction and torque on the normalized rotation radius $a/\xi$ and the rotation velocity $\omega\xi/c$.
The repetition rate of the steps increases with increasing rotation frequency $\omega\xi/c$. 
Furthermore, the number of observable jumps and their severity increase with increasing both the rotation frequency $\omega\xi/c$ and rotation radius. 

The Stoke's friction force vanishes at low velocities ($F \to 0$ as $V/c \to 0$) and scales linearly with $V$ at high velocities ($F \propto V/c$ as $V/c \to \infty$). 
In contrast, the friction torque approaches constant values in both the limits of slow and fast translational velocities.
The torque generally grows with an increase of $\omega a/c$.


\newpage
\section{Derivation of formula for Magnus force in the limit $ka\ll1$ and $\omega\gg kV$, $ck$}
The derivation starts with the expression given in the main text:
\begin{eqnarray}
{\bf F}_\textrm{M}=-i\sum_{k_1,k_2}({\bf k})_{\perp}X(k_1,k_2)\chi^{(2)}(k_1,k_2),
\end{eqnarray}
where $\mathbf{k} = \mathbf{k}_1 + \mathbf{k}_2$,
\begin{eqnarray}
X(k_1,k_2)&=&U^*(k_{12})U(k_1)U(k_2)\\
\nonumber
&=&
-(2\pi U_0)^3\sum_{n_1, n_2} J_{n_1+n_2}(k_{12} a) J_{n_1}(k_1 a) J_{n_2}(k_2 a)
e^{in_1(\phi_1-\phi_{12})}e^{in_2(\phi_2-\phi_{12})}\\
\nonumber
&&~~~~\times
\delta(\Omega_1 - \mathbf{k}_1\mathbf{V} - n_1\omega)\delta(\Omega_2 - \mathbf{k}_2\mathbf{V} - n_2\omega).
\end{eqnarray}
In the long-wavelength limit, $ka\ll1$, the dominant contribution arises from the terms with $n_1,n_2=0,\pm1$. The term corresponding to $n_1=n_2=0$ does not contribute to the non-dissipative anomalous transverse Magnus-like force. 

After straightforward algebraic manipulation, the remaining terms yield:
\begin{gather}
X(k_1,k_2)\approx-(2\pi U_0)^3\frac{a^2}{4}
\left\{({\bf k}_2\cdot{\bf k})-i[{\bf k}_2\times{\bf k}]_z\right\}
\delta(\Omega_1-{\bf k}_1\cdot{\bf V})\delta(\Omega_2-{\bf k}_2\cdot{\bf V}-\omega)
\\\nonumber
-(2\pi U_0)^3\frac{a^2}{4}
\left\{({\bf k}_2\cdot{\bf k})+i[{\bf k}_2\times{\bf k}]_z\right\}
\delta(\Omega_1-{\bf k}_1\cdot{\bf V})\delta(\Omega_2-{\bf k}_2\cdot{\bf V}+\omega)
\\\nonumber
-(2\pi U_0)^3\frac{a^2}{4}
\left\{({\bf k}_1\cdot{\bf k})-i[{\bf k}_1\times{\bf k}]_z\right\}
\delta(\Omega_1-{\bf k}_1\cdot{\bf V}-\omega)\delta(\Omega_2-{\bf k}_2\cdot{\bf V})
\\\nonumber
-(2\pi U_0)^3\frac{a^2}{4}
\left\{({\bf k}_1\cdot{\bf k})+i[{\bf k}_1\times{\bf k}]_z\right\}
\delta(\Omega_1-{\bf k}_1\cdot{\bf V}+\omega)\delta(\Omega_2-{\bf k}_2\cdot{\bf V}).
\end{gather}
Here, we used $J_0(x)\approx1,\,J_1(x)\approx x/2$ for small arguments, $x\ll1$.  
Substituting this relation into the expression for the Magnus force and utilizing the permutation $1\leftrightarrow 2$, gives
\begin{eqnarray}
{\bf F}_\textrm{M}&=&i(2\pi U_0)^3\frac{a^2}{2}\sum_{k_1,k_2}({\bf k})_{\perp}\chi^{(2)}(k_1,k_2)
\delta(\Omega_1-{\bf k}_1\cdot{\bf V})\\
\nonumber
&&\times
\Bigl[\left\{({\bf k}_2\cdot{\bf k})-i[{\bf k}_2\times{\bf k}]_z\right\}
\delta(\Omega_2-{\bf k}_2\cdot{\bf V}-\omega)
+\left\{({\bf k}_2\cdot{\bf k})+i[{\bf k}_2\times{\bf k}]_z\right\}
\delta(\Omega_2-{\bf k}_2\cdot{\bf V}+\omega)\Bigr].
\end{eqnarray}
Here, the scalar product ${\bf k}_2\cdot{\bf k}$ does not generate a transverse force and is therefore disregarded. 

Using the identity $[{\bf k}_2\times{\bf k}]_z=[{\bf k}_2\times{\bf k}_{1}]_z$, the Magnus force can be expressed as
\begin{gather}
{\bf F}_\textrm{M}=(2\pi U_0)^3\frac{a^2}{2}\sum_{k_1,k_2}({\bf k})_{\perp}[{\bf k}_1\times{\bf k}_{2}]_z\chi^{(2)}(k_1,k_2)
\delta(\Omega_1-{\bf k}_1\cdot{\bf V})
\Bigl[\delta(\Omega_2-{\bf k}_2\cdot{\bf V}-\omega)
-\delta(\Omega_2-{\bf k}_2\cdot{\bf V}+\omega)\Bigr].
\end{gather}
By exploiting the symmetry properties of the integrand under the transformations $k_1\rightarrow-k_1$ and $k_2\rightarrow-k_2$, it can be shown that only the real part of the response, $\text{Re}\,\chi^{(2)}(k_1,k_2)$, yields a finite contribution, ensuring that the force is real-valued. 
To cast the force relation into a symmetric form, we apply the permutation $k_1\leftrightarrow k_2$ followed by the substitution $k_1\rightarrow-k_1$, $k_2\rightarrow-k_2$ in the second term within the brackets, which yields:
\begin{eqnarray}
\label{EqForce02}
&&{\bf F}_\textrm{M}=(2\pi U_0)^3\frac{a^2}{2}\sum_{k_1,k_2}({\bf k}_{1}+{\bf k}_{2})_{\perp}[{\bf k}_1\times{\bf k}_{2}]_z\text{Re}\chi^{(2)}(k_1,k_2)\\
\nonumber
&&~~~~~~~~\times
\Bigl[\delta(\Omega_1-{\bf k}_1\cdot{\bf V})\delta(\Omega_2-{\bf k}_2\cdot{\bf V}-\omega)
-\delta(\Omega_1-{\bf k}_1\cdot{\bf V}-\omega)\delta(\Omega_2-{\bf k}_2\cdot{\bf V})\Bigr].
\end{eqnarray}
This relation explicitly demonstrates that the Magnus force vanishes in the absence of either rotation ($\omega=0$) or translational motion (${V}=0$), as expected. 

The expression for the real part of the second-order nonlinear susceptibility in the limit of infinitesimal damping ($\delta \to 0^+$), with frequencies replaced by wave numbers according to the delta-function resonance conditions, reads:
\begin{equation}
\text{Re}\left[\chi^{(2)}(k_1, k_2)\right] = \frac{n_c}{2m^2} \times \left( T_{\mathcal{P}} - \pi^2 \cdot c^2 \left[ T_{01} + T_{02} + T_{12} \right] \right),
\end{equation}
where
\begin{equation}
    T_{\mathcal{P}} = \mathcal{P} \left[ \frac{k^2 \Omega_1 \Omega_2 (\mathbf{k}_1 \cdot \mathbf{k}_2) + \Omega \Omega_1 k_2^2 (\mathbf{k} \cdot \mathbf{k}_1) + \Omega \Omega_2 k_1^2 (\mathbf{k} \cdot \mathbf{k}_2)}{\left(\Omega^2 - c^2 k^2\right) \left(\Omega_1^2 - c^2 k_1^2\right) \left(\Omega_2^2 - c^2 k_2^2\right)} \right]
\end{equation}
is a Principal value integral, with $\Omega = \Omega_1 + \Omega_2$;
\begin{equation}
    \begin{aligned}
    T_{01} = {} & \frac{\delta\left(\Omega^2 - c^2 k^2\right) \delta\left(\Omega_1^2 - c^2 k_1^2\right)}{\Omega_2^2 - c^2 k_2^2} \cdot k \cdot k_1 
    \left[ \left( k k_2 + k_1 k_2 + k_2^2 \right) (\mathbf{k}_1 \cdot \mathbf{k}_2) + k_1 k_2^2 (k_1 + k_2) \right]
    \end{aligned}
\end{equation}
is the first resonance contribution, evaluated under the delta-function conditions: $\Omega = c k$ and $\Omega_1 = c k_1$;
\begin{equation}
    \begin{aligned}
    T_{02} = {} & \frac{\delta\left(\Omega^2 - c^2 k^2\right) \delta\left(\Omega_2^2 - c^2 k_2^2\right)}{\Omega_1^2 - c^2 k_1^2} \cdot k \cdot k_2 
    \left[ \left( k k_1 + k_1 k_2 + k_1^2 \right) (\mathbf{k}_1 \cdot \mathbf{k}_2) + k_1^2 k_2 (k_1 + k_2) \right]
    \end{aligned}
\end{equation}
is the second resonance contribution for $\Omega = c k$ and $\Omega_2 = c k_2$;
\begin{equation}
    \begin{aligned}
    T_{12} = {} & \frac{\delta\left(\Omega_1^2 - c^2 k_1^2\right) \delta\left(\Omega_2^2 - c^2 k_2^2\right)}{\Omega^2 - c^2 k^2} \cdot k_1 \cdot k_2 
    \left[ 2 (\mathbf{k}_1 \cdot \mathbf{k}_2)^2 + \left( k_1^2 + k_2^2 + (k_1 + k_2)^2 \right) (\mathbf{k}_1 \cdot \mathbf{k}_2) + k_1 k_2 (k_1 + k_2)^2 \right]
    \end{aligned}
\end{equation}
is the third resonance contribution for $\Omega_1 = c k_1$, $\Omega_2 = c k_2$, and $\Omega = c(k_1 + k_2)$.



Furthermore, general expression~\eqref{EqForce02} can be simplified by exploiting the permutation symmetry $k_1\leftrightarrow k_2$. Consequently, the second term in the brackets doubles the contribution of the first, yielding:
\begin{eqnarray}
\label{HFLEqForce}
{\bf F}_\textrm{M}=(2\pi U_0)^3a^2\sum_{k_1,k_2}({\bf k}_{1}+{\bf k}_{2})_{\perp}[{\bf k}_1\times{\bf k}_{2}]_z\text{Re}\chi^{(2)}(k_1,k_2)
\delta(\Omega_1-{\bf k}_1\cdot{\bf V})\delta(\Omega_2-{\bf k}_2\cdot{\bf V}-\omega).
\end{eqnarray}
Analytical integration can be performed here in the case of fast rotation, $\omega\gg ck,Vk$. 
In this limit, the response functions can be simplified: 
\begin{equation}
\label{HFLEqResponse}
\text{Re}\chi^{(2)}(k_1, k_2)\approx \frac{n_c}{2m^2}\frac{\Omega_2^2k_1^2({\bf k}\cdot{\bf k}_2)}{\omega^4[\Omega^2_1 - c^2 k^2_1]}.
\end{equation}
Substituting this expression in Eq.~\eqref{HFLEqForce} and then, integrating over $\Omega_{1,2}$ gives:
\begin{eqnarray}
\label{HFLEqForce03}
{\bf F}_\textrm{M}=\frac{(2\pi U_0)^3a^2n_c}{2m^2\omega^4}\int\frac{d{\bf k}_1}{(2\pi)^2}\int\frac{d{\bf k}_2}{(2\pi)^2}
({\bf k}_{1}+{\bf k}_{2})_{\perp}[{\bf k}_1\times{\bf k}_{2}]_z
\frac{(\omega+{\bf k}_2\cdot{\bf V})^2k_1^2({\bf k}\cdot{\bf k}_2)}{({\bf k}_1\cdot{\bf V})^2 - c^2 k^2_1},
\end{eqnarray}
where the integral is taken as a principal value and, as before,  $\mathbf{k} = \mathbf{k}_1 + \mathbf{k}_2$.

We can evaluate these 2D wave vector integrals in cylindrical system of coordinates, placing the upper integration limit by $k_0\sim 1/\xi$. 
Without the loss of generality, we take the velocity directed along the x-axis, such that $\mathbf{V} = (V, 0)$. Then, the transverse ($y$) component of the force is given by:
\begin{eqnarray}
F_{\textrm{M}y} = \frac{(2\pi U_0)^3a^2n_c}{2m^2\omega^4} \int\limits_0^{k_0}
\int\limits_0^{2\pi} \frac{k_1dk_1d\phi_1}{(2\pi)^2} 
\int\limits_0^{k_0}
\int\limits_0^{2\pi}
\frac{k_2dk_2d\phi_2}{(2\pi)^2} 
(k_{1y} + k_{2y}) [\mathbf{k}_1 \times \mathbf{k}_2]_z
\frac{(\omega + \mathbf{k}_2 \cdot \mathbf{V})^2 k_1^2 (\mathbf{k} \cdot \mathbf{k}_2)}{(\mathbf{k}_1 \cdot \mathbf{V})^2 - c^2 k_1^2}   
\end{eqnarray}
in polar coordinates, $\mathbf{k}_1 = (k_1, \phi_1)$ and $\mathbf{k}_2 = (k_2, \phi_2)$, thus the velocity dot products become $\mathbf{k}_1 \cdot \mathbf{V} = k_1 V \cos\phi_1$ and $\mathbf{k}_2 \cdot \mathbf{V} = k_2 V \cos\phi_2$.

After straightforward algebraic calculations, we find:
\begin{eqnarray}
F_{\textrm{M}y} = \frac{\pi U_0^3 a^2 n_c k_0^{10}}{96 m^2 \omega^3 V} \left[ 6 - \frac{6c^2 - 5V^2}{c\sqrt{c^2 - V^2}}\theta[c-V] \right].    
\end{eqnarray}
%
To evaluate the behavior of the force as $V \to 0$, we perform a Taylor expansion on the term inside the brackets, assuming 
$V/c \ll 1$, that yields
\begin{eqnarray}
F_{\textrm{M}y}(V \to 0) \approx \frac{\pi U_0^3 a^2 n_c k_0^{10}}{96 m^2 \omega^3 V} \left[ 2 \frac{V^2}{c^2} \right]    
= 
\frac{\pi U_0^3 a^2 n_c k_0^{10}}{48 m^2 \omega^3 c^2} V.
\end{eqnarray}
Therefore, the force scales linearly with velocity at low speeds, manifesting an effective viscous drag or transverse drift friction.




\bibliography{biblio}
\bibliographystyle{apsrev4-2}
